# A new electromagnetic code for ICRF antenna in EAST[*]

YANG Hua(杨桦)[1]   WU Cong-Feng(吴丛凤)[1;1)]   DONG Sai(董赛)[1]   ZHANG Xin-Jun(张新军)[2]   ZHAO Yan-Ping(赵燕平)[2]   SHANG Lei(尚雷)[1]

[1] National Synchrotron Radiation Laboratory, University of Science and Technology of China, Hefei 230029, China

[2] Institute of Plasma Physics, Chinese Academy of Sciences, Hefei 230031, China

**Abstract:** The demand for an effective tool to help in the design of ion cyclotron radio frequency (ICRF) antenna system for fusion experiment has driven the development of predictive codes. A new electromagnetic code based on the method of moments (MOM) is described in the paper. The code computes the electromagnetic field by the solution of the electric field integral equation. The structure of ICRF antennas are discretized with triangular mesh. By using the new code, the scattering parameter and the surface current are given and compared with the result by commercial code CST. Moreover, the power spectra are studied with different toroidal phases for heating and current drive. Good agreement of simulation results between the new code and CST are obtained. The code has been validated against CST for EAST ICRF antenna.

**Key words:** ICRF antenna, electromagnetic field, code, EAST

**PACS:** 52.50.Qt, 52.40.Fd

## 1 Introduction

Radio frequency wave in the ICRF range is an important heating method in Tokamak. It can provide heating, current drive and other physics research. ICRF antenna is a key component in ICRF system. The performance of antenna is limited by arcing, impurities and current density [1]. A proper prediction of the antenna coupling properties is an essential ingredient in estimating the performance of future ICRF antenna systems. The coupling properties of antenna sensitively depend on near-fields, self-consistent surface currents on straps, plasma density profile, the spectrum and so on.

The performance forecast of ICRF antenna has driven the development of code such as ICANT [2], RANT3D [3], FELICE [4] and ARGUS [5]. In recent years, commercial codes, like HFSS [6] and CST Microwave Studio (MWS) [7], which have a good ability to hand complex antenna geometry, were used for ICRF antenna performance analysis. However, they currently do not feature any model of magnetized plasma. This is a serious limitation to the application of these codes to the magnetic fusion environment.

We wanted to achieve the same capabilities of geometrical modelling as commonly achieved in commercial codes for the analysis of 'conventional' microwave antennas in the code. Moreover, the magnetized, inhomogeneous and hot plasma in front of the antenna are expected to be taken into account in the future. Hence, the aim of our quest was to have an accurate description of the antenna structure and the fusion plasma. This paper is aimed at describing the substantive evolution that has occurred up to now in the theoretical framework of the new code, its implementation and the resulting status of the code.

The paper is organized as follows: first in section 2, the theory in the code about the electromagnetic field and the surface current calculation are stated. Then in section 3, by using the new code the scattering parameter and the surface current are shown and compared with the result obtained from commercial code CST [8]. Besides, the spectrum is presented. In section 4 the conclusions are given.

---

[*] Supported by the National Magnetic confinement Fusion Science Program (2010GB110000)

1) E-mail: cfwu@ustc.edu.cn

## 2 Theory

### 2.1 Antenna modeling

The accurate model of antenna geometry has significant effect on the calculation of electromagnetic field. There are two methods of creating the antenna structure in MATLAB. One of methods is the use of the partial differential equation (PDE) toolbox. Graphical user interface (GUI), which is convenient to mesh object, is employed in PDE toolbox. The equilateral triangle and right triangle are both made in the toolbox. For example, a rectangle with 2 m width and 1 m length is meshed by two ways in Fig. 1. The other method is to use the Delaunay function. It can be used for all objects. In the code, both methods may be chosen to mesh an antenna structure. The object will be discretized with triangular mesh which is saved for next step.

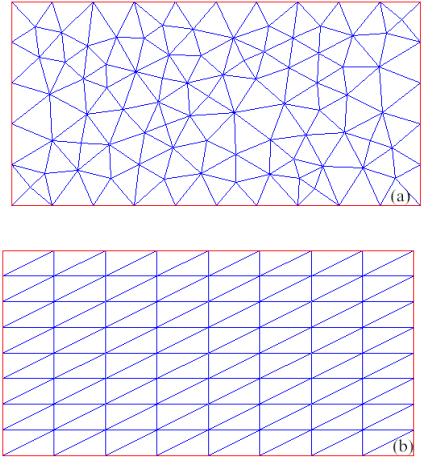

Fig.1. The equilateral triangle (a) mesh and right triangle (b) mesh for a rectangle

### 2.2 The equation of MOM

The performance predictive aim of ICRF antenna is about the calculation of admittance matrix, current, spectrum and near field with complex antenna, which may contain straps, power feed structure and box and so on. In the code, the core task is the solution of the Maxwell equations. First, the electric field integral equation (EFIE) is to be given briefly [9, 10].

As the area of an open or closed perfectly conductor is S, the incident electric field $E^i$, which is assumed it is dependent $e^{j\omega t}$, is around the conductor. The scattered electric field $E^s$ can be derived by

$$E^s = -j\omega A - \nabla\Phi, \quad (1)$$

with the magnetic vector potential and scalar potential defined as

$$A(r) = \frac{\mu}{4\pi}\int_S J \frac{e^{-jkR}}{R}dS', \quad (2)$$

$$\Phi(r) = \frac{1}{4\pi\varepsilon}\int_S \sigma \frac{e^{-jkR}}{R}dS', \quad (3)$$

the charge density $\sigma$ and the current density $J$ are related though the charge continuity equation

$$\nabla_S \cdot J = -j\omega\sigma. \quad (4)$$

When the scatter field is being, the boundary condition is $(E^s + E^i)_{\tan} = 0$, so the scatter field

$$E^S_{\tan} = (-j\omega A - \nabla\Phi)_{\tan}. \quad (5)$$

The Eq. (2)-(5) is called electric field integral equation.

Then, the EFIE is to be expressed as a term of basis function. In Fig. 2 shows a pair triangle mesh. The coordinate of vertex of a triangle and sequential coding are created. The basis function is to be associated with the two triangles attached to non-boundary edge (interior edge). $T_n^+$ and $T_n^-$ are corresponding to the $n$th edge of triangle. The position vector $\rho_n^+$ is defined with respect to the free vertex of $T_n^+$. Similar remarks apply to the position vector $\rho_n^-$ except that it is directed towards the free vertex of triangle $T_n^-$. The method was proposed by Rao-Wilton-Glisson (RWG) [10]. The RWG basis function associated with the $n$th edge is formulated by

$$f_n(r) = \begin{cases} \dfrac{l_n}{2A_n^+}\rho_n^+, & r \text{ in } T_n^+ \\ \dfrac{l_n}{2A_n^-}\rho_n^-, & r \text{ in } T_n^- \\ 0, & \text{otherwise} \end{cases} \quad (6)$$

where $l_n$ is the length of the $n$th edge and $A_n^\pm$ is the area of triangle $T_n^\pm$.

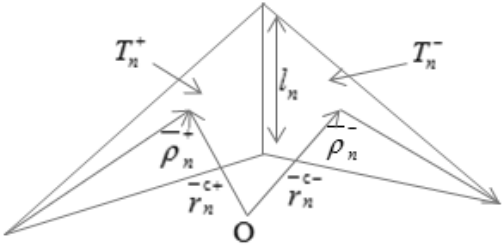

Fig.2. A pair of triangle and interior edge

The surface divergence of $f_n$ can be expressed as

$$\nabla_S \cdot f_n(r) = \begin{cases} \dfrac{l_n}{A_n^+}, & r \text{ in } T_n^+ \\ -\dfrac{l_n}{A_n^-}, & r \text{ in } T_n^- \\ 0, & \text{otherwise} \end{cases} \quad (7)$$

The current on S may be expressed in terms of the $f_n$ as

$$J = \sum_{n=1}^{N} I_n f_n(r). \quad (8)$$

It is noted that each coefficient $I_n$ may be interpreted as the component of current density flowing past the $n$th edge.

The EFIE (5) can be written as

$$\langle E^i, f_m \rangle = j\omega \langle A, f_m \rangle + \langle \nabla\Phi, f_m \rangle. \quad (9)$$

With Eq. (6), the left in Eq. (9) may be approximated as follows

$$\langle E^i, f_m \rangle$$
$$= \frac{l_m}{2A_m^+}\int_{T_m^+} E^i \cdot \rho_m^+ ds + \frac{l_m}{2A_m^-}\int_{T_m^-} E^i \cdot \rho_m^- ds$$
$$\cong \frac{l_m}{2} E^i(r_m^{c+}) \cdot \rho_m^+ + \frac{l_m}{2} E^i(r_m^{c-}) \cdot \rho_m^-. \quad (10)$$

With similar approximations, the vector potential in Eq. (9) is written as

$$\langle A, f_m \rangle = \frac{l_m}{2A_m^+}\int_{T_m^+} A \cdot \rho_m^+ ds + \frac{l_m}{2A_m^-}\int_{T_m^-} A \cdot \rho_m^- ds$$
$$\cong \frac{l_m}{2} A(r_m^{c+}) \cdot \rho_m^+ + \frac{l_m}{2} A(r_m^{c-}) \cdot \rho_m^-. \quad (11)$$

The last item of the formula (9) can be expressed as

$$\langle \nabla\Phi, f_m \rangle = -\int_S \Phi \nabla_S \cdot f_m ds$$
$$= -\frac{l_m}{A_m^+}\int_{T_m^+}\Phi ds + \frac{l_m}{2A_m^-}\int_{T_m^-}\Phi ds$$
$$\cong -l_m \Phi(r_m^{c+}) + l_m \Phi(r_m^{c-}). \quad (12)$$

With Eq. (10)-(12), the Eq. (9) now becomes

$$l_m[E^i(r_m^{c+}) \cdot \frac{\rho_m^+}{2} + E^i(r_m^{c-}) \cdot \frac{\rho_m^-}{2}]$$
$$= j\omega l_m[A(r_m^{c+}) \cdot \frac{\rho_m^+}{2} + A(r_m^{c-}) \cdot \frac{\rho_m^-}{2}]$$
$$+ l_m[\Phi(r_m^{c-}) - \Phi(r_m^{c+})]$$
$$+ l_m[\Phi(r_m^{c-}) - \Phi(r_m^{c+})]. \quad (13)$$

Substitution of the formula (2), (3) and (8) into Eq. (13) yields an expression as

$$l_m[E^i(r_m^{c+}) \cdot \frac{\rho_m^+}{2} + E^i(r_m^{c-}) \cdot \frac{\rho_m^-}{2}]$$
$$= l_m[j\omega\frac{\mu}{4\pi}(\int_S I_n f_n(r') \frac{e^{-jkR_m^+}}{R_m^+}ds' \cdot \frac{\rho_m^+}{2}$$
$$+ \int_S I_n f_n(r') \frac{e^{-jkR_m^-}}{R_m^+}ds' \cdot \frac{\rho_m^-}{2})]$$
$$+ l_m[\frac{1}{4\pi\varepsilon}(\int_S \frac{I_n \nabla_S' \cdot f_n(r')}{-j\omega} \frac{e^{-jkR_m^-}}{R_m^+}ds'$$
$$- \int_S \frac{I_n \nabla_S' \cdot f_n(r')}{-j\omega} \frac{e^{-jkR_m^+}}{R_m^+}ds')]. \quad (14)$$

Assume that

$$V_m = l_m[E^i(r_m^{c+}) \cdot \frac{\rho_m^+}{2} + E^i(r_m^{c-}) \cdot \frac{\rho_m^-}{2}], \quad (15)$$

$$A_{mn}^\pm = \frac{\mu}{4\pi}\int_S f_n(r')\frac{e^{-jkR_m^\pm}}{R_m^\pm}ds', \quad (16)$$

$$\Phi_{mn}^\pm = -\frac{1}{4\pi j\omega\varepsilon}\int_S \nabla_S' \cdot f_n(r')\frac{e^{-jkR_m^\pm}}{R_m^\pm}ds', \quad (17)$$

$$Z_{mn} = l_m[j\omega(A_{mn}^+ \cdot \frac{\rho_m^+}{2} + A_{mn}^- \cdot \frac{\rho_m^-}{2}) \\ + \Phi_{mn}^- - \Phi_{mn}^+], \quad (18)$$

$$R_m^\pm = |r_m^{c\pm} - r'|. \quad (19)$$

The formula (14) can be written compactly as
$$Z_{mn}I_n = V_m. \quad (20)$$

The formula (20) is called the moment equation. Once the forcing vector $V$ is determined and the elements of $Z$ is calculated by Eq. (18), the vector $I_n$ can be derived. If the phases of feeding voltage for each port are not coincident, the feeding voltage will be amended with
$$V_m = V_m \times \exp(j \times \text{phase}). \quad (21)$$

**2.3 Electromagnetic field and the post processing**

The aim of this section is to introduce how to calculate the electromagnetic field, which within the antenna and around it at a given input power level is needed to check the potential danger of rectification and other adverse nonlinear effects [11]. The solution method of field and post processing are given in the following paper.

If the vector $I_n$ has been known, the electromagnetic field in any point is also easily calculated [12]. The current on RWG edge is the equivalent of one electric dipole. The dipole moment $m$ may be expressed as
$$m = \int_S J ds = \int_S I_n f_n ds$$
$$= I_n \int_S f_n ds = l_n I_n (r_n^{c-} - r_n^{c+}). \quad (22)$$

With Eq. (22), the electromagnetic field expression at $r$ in the distance can be deduce by the dipole in the origin as follow
$$H(r) = \frac{jk}{4\pi r^2}(m \times r)(1 + \frac{1}{jkr})e^{-jkr}, \quad (23)$$
$$E(r) = \frac{\eta}{4\pi r}[(\frac{(r \bullet m)r}{r^2} - m)(jk + \frac{1}{r} + \frac{1}{jkr^2})$$
$$+ 2\frac{(r \bullet m)r}{r^2}(\frac{1}{r} + \frac{1}{jkr^2})]e^{-jkr}. \quad (24)$$

The Poynting vector also can be written as
$$W(r) = \frac{1}{2}\text{Re}[E(r) \times H^*(r)]. \quad (25)$$

In the post processing, the field and Poynting vector along a line or distributed in a face can be easily acquired with above formulae.

# 3 Result and discussion

In order to validate the reliability and accuracy of the code, a series of parameters based on the double strap antenna in Experimental Advanced Superconducting Tokamak (EAST) are compared with the code and CST MWS. The antenna model in CST MWS is shown in Fig. 3(a). The model contains current straps, box, feed port and ground structure, where x, y, z correspond to the radial, poloidal and toroidal coordinates. The antenna operates with 6 MW during 30-110MHz [13]. The length and width of straps are respectively 540 mm and 110 mm [14].

Fig. 3(b) shows the model and triangular mesh of the antenna by the code. This triangular mesh is suitable for arbitrarily shaped objects. The voltage across a narrow gap between the feeder and a back-wall represents the power feed. The number of RWG edges and mesh of triangles in the structure are 1894 and 1976. There is more than one mesh across the strap that means the code can simulate the current distribution along the strap transverse.

For each frequency point model takes less than 20 s to execute on a Windows7 base PC with a 2.8 GHz and 4 GB RAM. Among the total time, the element of $Z$ matrix in Eq. (20) is filled with about 7.87 s and the solution of Eq. (20) spent approximately 2.17 s.

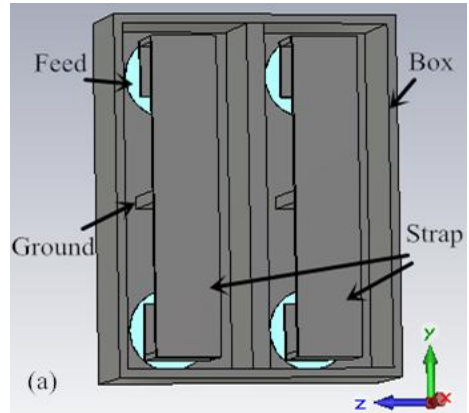

Fig.3 (a) Model of the double loop ICRF antenna in EAST is created by CST MWS. The strap is grounded in the center and the RF power is fed from the end of the strap.

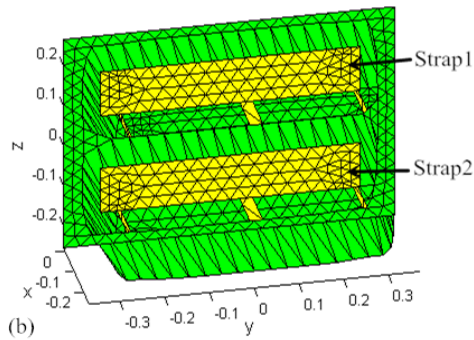

Fig.3 (b) Model and triangular mesh of antenna is simulated by the code. There are 1944 triangular.

### 3.1 validation of code

The antenna scattering matrix (S-matrix) is of crucial importance to assess compliance with power handling requirements and to design the tuning-and-matching system. Firstly, in order to verify the rationality of the model in CST MWS, the S11 in port 1 is calculated with water instead of plasma that is a common method. The S11 in Fig. 4 is reasonable [15]. So the antenna model will be used to check the new code.

With the wave propagation in vacuum, the S11 are compared with the data derived from CST MWS and the new code in Fig. 5. The agreement between the two codes is in the range of 0-500MHz except 206MHz which may be interpreted by the mesh of model is not enough. The antenna material is assumed a perfect electric conductor. The resonance frequency is about 410MHz.

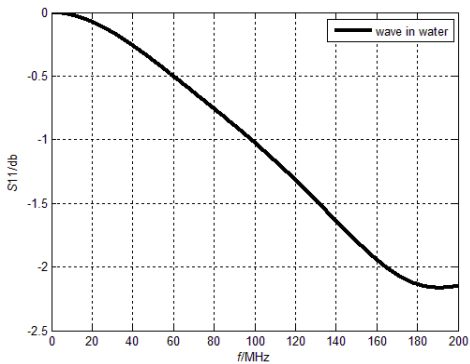

Fig.4 The scattering parameter S11 in port 1 with loading water is shown.

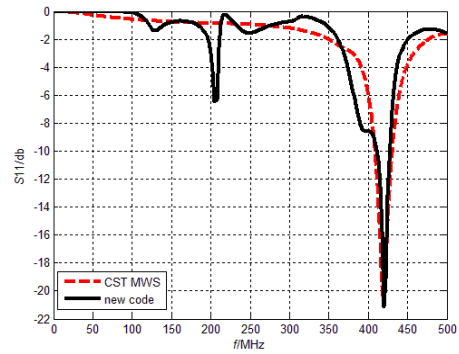

Fig.5 The parameter S11 is computed by CST MWS (the dotted line) and the new code (The solid line).

### 3.2 The distribution of surface current

This code is able to provide important information on the RF surface current and electric field distributions as well as the useful CST MWS code. The surface current on the antenna is checked for whether the current is self-consistent. The surface current distribution is shown in Fig. 6. In Fig. 6(a) and (b), the high density current is both at the location of the strap edge, feed port and grounding position. The current is self-consistent. The toroidal current phasing is monopole (0 0) phasing.

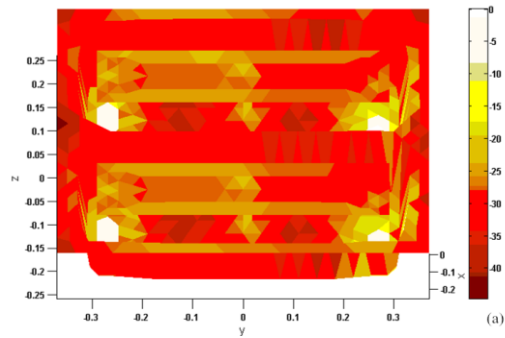

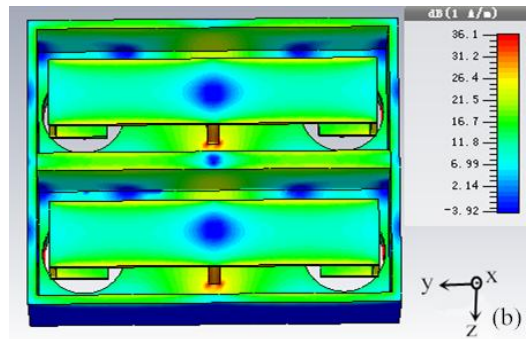

Fig.6. (a) The current distribution is calculated by the new code (a) and CST MWS (b).

### 3.3 Power spectra

Power spectra are needed to evaluate the antenna ability to effectively coupling power to the plasma because of the wave propagation and dissipation in plasma is related to power spectrum [16]. With the new code, the near field is calculated by section 2.3 equations. Then, the power is expressed as a relationship with $k_z$ by Fourier transform. The result is shown in Fig. 7 with different toroidal phasing for heating (0 π) and current drive (0 π/2). In the figure, the amplitude of the power is a relative value because of it is related to the feed power. The power spectrum is symmetric for heating phase and asymmetric for current drive phase. The result is also consistent with the Ref. [17]. So the calculation result is reasonable.

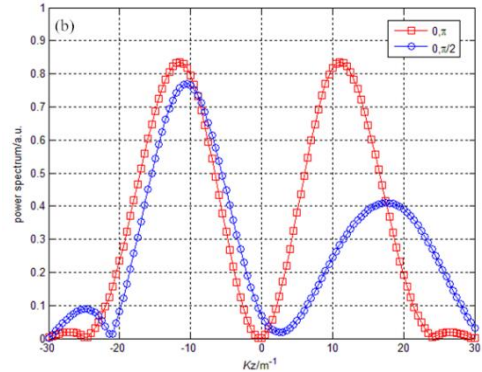

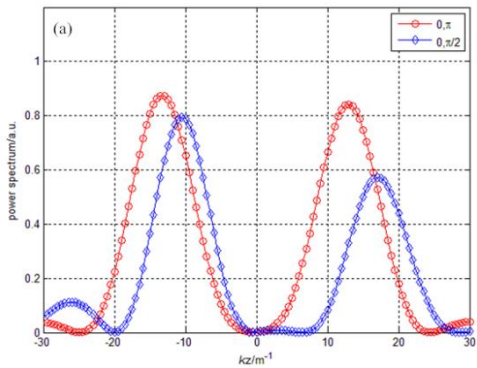

Fig.7. The power spectra with toroidal phasing (0 π) and (0 π/2). The data are acquired from the new code (a) and CST MWS (b).

### 4 Conclusions

The theory of the new code with MOM has been described. The solution of Maxwell's equations is by the solution of the EFIE. The mesh of antenna structure is used with triangular mesh which is suitable for arbitrarily shaped objects. The performance of the EAST double loops antenna is evaluated by the new code and CST MWS. The scattering parameter S11 in vacuum and surface current, which are agree with the result by CST MWS, are simulated. Besides, the power spectra deduced from near field are shown with two cases. From this analysis we conclude that the code is a useful tool for ICRF antenna. It is in progress that the fusion plasma is taken into account in the new code.